\pdfoutput=1

\documentclass[11pt,twoside,a4paper,cmspaper,final,collab]{cms-tdr}
\def\svnVersion{f725b7d-D}\def\svnDate{2020/08/21}\def\cmsCernNoTag{CERN-EP-2019-247}\def\cmsCernDate{\today}\def\cmsMessage{"Published in Physical Review Letters as \href{http://dx.doi.org/10.1103/PhysRevLett.125.102001}{\doi{10.1103/PhysRevLett.125.102001}.}"}

\begin{document}\cmsNoteHeader{HIN-18-007}

\hyphenation{had-ron-i-za-tion}
\hyphenation{cal-or-i-me-ter}
\hyphenation{de-vices}
\RCS$HeadURL$
\RCS$Id$

\newlength\cmsFigWidth
\ifthenelse{\boolean{cms@external}}{\setlength\cmsFigWidth{0.49\textwidth}}{\setlength\cmsFigWidth{0.65\textwidth}} 
\ifthenelse{\boolean{cms@external}}{\providecommand{\cmsLeft}{upper\xspace}}{\providecommand{\cmsLeft}{left\xspace}}
\ifthenelse{\boolean{cms@external}}{\providecommand{\cmsRight}{lower\xspace}}{\providecommand{\cmsRight}{right\xspace}}
\ifthenelse{\boolean{cms@external}}{\providecommand{\suppMaterial}{the supplemental material~\cite{supp}}}{\providecommand{\suppMaterial}{Appendix~\ref{app:suppMat}}}

\cmsNoteHeader{HIN-18-007} 

\newcommand{\Dzero}{\PDz}
\newcommand {\ptj}    {\ensuremath{\pt^\mathrm{jet}}\xspace}
\newcommand {\etaj}    {\ensuremath{\eta^\mathrm{jet}}\xspace}
\newcommand {\ptd}    {\ensuremath{\pt^\mathrm{D}}\xspace}
\newcommand {\Njd}    {\ensuremath{N_\mathrm{jD}}\xspace}
\newcommand {\ak}    {anti-\kt}
\newcommand {\sqrtsnn}  {\sqrtsNN}
\newcommand{\sqrts} {\ensuremath{\sqrt{\smash[b]{s}}}\xspace}
\newcommand{\pp} {\Pp{}\Pp}
\newcommand{\PbPb} {PbPb\xspace}

\title{Studies of charm quark diffusion inside jets using \texorpdfstring{\PbPb and \pp collisions at $\sqrtsNN=5.02\TeV$}{PbPb and pp collisions at sqrt(s[NN]) = 5.02 TeV}}

\date{\today}

\abstract{
The first study of charm quark diffusion with respect to the jet axis in heavy ion collisions is presented. The measurement is performed using jets with $\ptj> 60\GeVc$ and \Dzero mesons with $\ptd>4\GeVc$ in lead-lead (\PbPb) and proton-proton (\pp) collisions at a nucleon-nucleon center-of-mass energy of $\sqrtsNN=5.02\TeV$, recorded by the CMS detector at the LHC. The radial distribution of \Dzero mesons with respect to the jet axis is sensitive to the production mechanisms of the meson, as well as to the energy loss and diffusion processes undergone by its parent parton inside the strongly interacting medium produced in \PbPb collisions. When compared to Monte Carlo event generators, the radial distribution in \pp collisions is found to be well-described by \PYTHIA, while the slope of the distribution predicted by \SHERPA is steeper than that of the data. In \PbPb collisions, compared to the \pp results, the \Dzero meson distribution for $4<\ptd<20\GeVc$ hints at a larger distance on average with respect to the jet axis, reflecting a diffusion of charm quarks in the medium created in heavy ion collisions. At higher \ptd, the \PbPb and \pp radial distributions are found to be similar.
}

\hypersetup{
pdfauthor={CMS Collaboration},
pdftitle={Measurement of the radial profile of D0 mesons in jets produced in pp and PbPb collisions at 5.02 TeV},
pdfsubject={CMS},
pdfkeywords={CMS, physics, relativistic nuclear collisions, jets}}

\maketitle 

The quark gluon plasma (QGP), the deconfined matter created in collisions of heavy ions accelerated to ultrarelativistic energies~\cite{Shuryak:1978ij, Shuryak:2004cy}, can be probed by studying the remnants of hard scatterings occurring in this medium. The outgoing partons (quarks and gluons), which produce final-state jets of particles, interact strongly with the QGP and lose energy~\cite{Appel:1985dq, d'Enterria:2009am,Wiedemann:2009sh}, a phenomenon known as jet quenching, as observed at the BNL RHIC~\cite{PhysRevLett.97.162301,PhysRevC.77.011901} and the CERN LHC~\cite{Aad:2010bu, Chatrchyan:2011sx, Adam20151}. Jet quenching results in modifications of the energy and structure of jets observed in heavy-ion collisions, compared to proton-proton (\pp) collisions. One of the most striking features of jet quenching is the enhanced production of low transverse momentum hadrons ($\pt\approx 2\text{--}5\GeVc$) at large angles with respect to the final-state jet axis. This phenomenon manifests itself in the form of modifications of the jet fragmentation function~\cite{Chatrchyan:2014ava, Aaboud:2017bzv,Sirunyan:2018qec}, as well as the jet radial profile and the energy flow~\cite{Chatrchyan:2013kwa, Khachatryan:2015lha, Sirunyan:2018jqr,Sirunyan:2018ncy}. Interpretations of experimental results include medium-induced gluon radiation, modification of jet splitting functions, and medium response to the hard scattered partons~\cite{CasalderreySolana:2007zz, d'Enterria:2009am, Wiedemann:2009sh, Majumder:2010qh, Burke:2013yra}. 

Studying heavy flavor (HF) mesons in jets should give further insight into the origin of the observed modifications for light flavored particles~\cite{Wang:2019xey} and can provide new information about HF jet fragmentation in both \pp and lead-lead (\PbPb) collisions. Moreover, measurements of angular correlations between HF mesons and jets can be used to constrain parton energy loss mechanisms and to better understand the heavy-quark diffusion (\ie, propagation) inside the medium~\cite{Nahrgang:2013saa,Cao:2015cba,Hambrock:2017sno,Wang:2019xey,Dong:2019byy}. This is complementary information to that obtained with measurements of inclusive HF meson spectra~\cite{Adam:2015sza,Adam:2015nna,Sirunyan:2017xss,Sirunyan:2017oug,Khachatryan:2016ypw}, HF meson azimuthal anisotropy~\cite{Abelev:2014ipa,Khachatryan:2016ypw,Sirunyan:2017plt,Adamczyk:2017xur,Acharya:2017qps}, and HF-tagged jets~\cite{Chatrchyan:2013exa,Sirunyan:2018jju}.

In this Letter, the first measurements of the radial distributions of \Dzero mesons in jets from the same parton scattering are presented, for two \Dzero meson \pt intervals: a low-\pt interval $4<\ptd<20\GeVc$, and a high-\pt one, $\ptd>20\GeVc$. The \Dzero mesons are measured via their hadronic decay channels $ \PDz\to \PKm \PGpp$ and $\PADz \to \PKp \PGpm$ with the CMS detector at the LHC. The observable is the normalized radial distribution of the \Dzero meson with respect to the jet axis, defined as 
\begin{linenomath}
\begin{equation}
\frac{1}{\Njd}\frac{\rd\Njd}{\rd r} = \frac{1}{\Njd}\frac{\left.\Njd\right|_{\Delta r}}{\Delta r},
\label{eq:dNdr}
\end{equation}
\end{linenomath}
where the distance from the jet axis, $r=\sqrt{\smash[b]{(\Delta\phi_{\mathrm{jD}})^{2}+(\Delta\eta_{\mathrm{jD}})^{2}}}$, is defined as the quadratic sum of the differences in pseudorapidity\,($\Delta\eta_{\mathrm{jD}}$) and azimuth\,($\Delta\phi_{\mathrm{jD}}$) of the \Dzero meson with respect to the jet axis, and $\Delta r$ is the width of the $r$ interval. The quantity $\left.\Njd\right|_{\Delta r}$ is the number of \Dzero mesons in the $\Delta r$ interval, and \Njd is the integral of the distribution in the $r$ region from 0 to 0.3, the distance parameter used for the jet reconstruction. 

The main feature of the CMS detector~\cite{Chatrchyan:2008zzk} is a superconducting solenoid, providing a magnetic field of 3.8\unit{T}. Within the solenoid volume is a silicon pixel and strip tracker, which is used to detect charged particles, a lead tungstate crystal electromagnetic calorimeter, and a brass and scintillator hadron calorimeter, each composed of a barrel and two endcap sections. Hadron forward calorimeters extend the coverage up to $\abs{\eta}=5.2$ and are used for collision event selection. Muons are detected in gas-ionization chambers embedded in the steel flux-return yoke outside the solenoid.

The \pp (\PbPb) data set used in this analysis corresponds to an integrated luminosity of 27.4\pbinv (404\mubinv). High-\pt jet events were selected by a high-level trigger algorithm~\cite{Khachatryan:2016bia} with a \ptj threshold of 60\GeVc. For the offline analysis, events must pass a set of selection criteria designed to reject beam-gas collisions and beam scraping events~\cite{Khachatryan:2016odn,Khachatryan:2010xs}. The \PbPb results are reported for the inclusive sample: no selection on centrality (\ie, the degree of overlap of the two colliding nuclei) is made.

Several Monte Carlo (MC) simulated event samples are used to evaluate the background contributions, signal efficiencies, and detector acceptance corrections. The simulated events include both prompt (produced directly from the {\cPqc} quark fragmentation) and nonprompt (from {\cPqb} hadron decays) \PDz{} meson events. The \pp collisions are generated using \PYTHIA v.8.212~\cite{Sjostrand:2014zea}, tune CUETP8M1~\cite{Khachatryan:2015pea}. The \EVTGEN~1.3.0~\cite{evtgen} generator is used to simulate \Dzero meson and {\cPqb} hadron decays, and final-state photon radiation in the \PDz{} meson decays is simulated with \PHOTOS 2.0~\cite{Barberio:1990ms}. For the \PbPb MC samples, each \PYTHIA event is embedded into a \PbPb collision event generated with \HYDJET~1.9~\cite{Lokhtin:2005px}, which is tuned to reproduce global event properties. The MC events are propagated through the CMS detector using the \GEANTfour package~\cite{geant4}.

The particle-flow (PF) algorithm~\cite{CMS-PRF-14-001} is used to reconstruct and identify each individual particle in a \pp or \PbPb event. To form jets, the PF particles are clustered using an \ak algorithm provided by the {\sc{FastJet}} framework~\cite{Cacciari:2008gp,Cacciari:2011ma} with a distance parameter of 0.3. In order to subtract the underlying event (UE) background in \PbPb collisions~\cite{Chatrchyan:2011sx,Chatrchyan:2012nia}, an iterative algorithm~\cite{Kodolova:2007hd} is employed. In \pp collisions jets are reconstructed without UE subtraction. The jet energy corrections are derived from simulation, separately for \pp and \PbPb data, and are confirmed via energy-balance methods applied to dijet, multijet, photon+jet, and leptonically decaying $\PZ$+jet events in \pp data~\cite{Khachatryan:2016kdb}. Jets with $\abs{\etaj} < 1.6$ and corrected $\ptj > 60\GeVc$ are selected for this analysis.

The \PDz{} candidates are reconstructed by combining pairs of oppositely-charged particle tracks with an invariant mass within $\pm$0.2\GeVcc of the world-average \PDz{} meson mass, 1.8\GeVcc~\cite{pdg:2016}. They are reconstructed independently from the PF jets, which do use the same track collection. In order to suppress the combinatorial background, each track is required to have $\pt>2\GeVc$, to be within $\abs{\eta} < 2$, and pass a set of quality selections~\cite{Khachatryan:2016odn}. For each pair of selected tracks, two \PDz{} candidates are created by assuming that one of the particles has the mass of the pion while the other has the mass of the kaon, and vice-versa. The \PDz{} candidates are required to have rapidity $\abs{y}< 2$ and $\ptd>4\GeVc$. They are further paired with every selected jet in the same event, and have their invariant mass distributions recorded in two \ptd bins, $4<\ptd<20\GeVc$ and $\ptd>20\GeVc$, and four $r$ bins, 0--0.05, 0.05--0.1, 0.1--0.3, and 0.3--0.5. In order to reduce further the combinatorial background, the \PDz{} candidates are required to pass three additional topological selections. The three-dimensional (3D) decay length (distance between the primary vertex and \Dzero secondary vertex ${L_\mathrm{3D}}$) normalized to its uncertainty is required to be larger than 2.34--4.00. The pointing angle $\theta_\mathrm{p}$ (defined as the angle between the total momentum vector of the \Dzero candidate and the vector connecting the primary and secondary vertices) is required to be smaller than 0.020--0.046 radians. In both cases (the 3D decay length and $\theta_\mathrm{p}$), the selection criteria depend on the \ptd and $r$ bins, and are optimized separately for the \pp and \PbPb data. The selection is optimized using a multivariate technique~\cite{TMVA} in order to maximize the statistical significance of the \PDz{} meson signals. Tighter selections are found for the low-\ptd bin, with increasing or decreasing $r$ values, for $\theta_\mathrm{p}$ and the 3D decay length significance, respectively. Finally, the $\chi^{2}$ probability of the secondary vertex fit is required to be larger than 5\%. These selections ensure a prompt \PDz{} meson fraction larger than 80\% in both \ptd bins of this analysis. 

The \PDz{} meson yield in each \pt and $r$ interval is extracted with a binned maximum likelihood fit to the invariant mass distributions in the range $1.7 <m_{\PGp \PK}< 2.0\GeVcc$. The combinatorial background originating from random pairs of tracks not produced by a \PDz{} meson decay is modeled by a third-order polynomial. The signal shape is found to be best modeled by the sum of two Gaussian functions with the same mean but different widths. The two Gaussians are found to best capture the many contributions to the \PDz{} peak resolution from tracks with a highly $\eta$-dependent \pt resolution. The common mean of the Gaussian functions, the \Dzero yield, and all the background parameters are free parameters in the fit. An additional Gaussian function with a larger width is used to describe the invariant mass shape of \PDz{} candidates with an incorrect mass assignment from the exchange of the pion and kaon designations. The widths of the Gaussian functions that describe the \PDz{} signal shape and the shape of the \PDz{} candidates with swapped mass assignments are fixed by simulation, after correcting for the difference in resolution between data and MC. The ratio between the numbers of the signal \PDz{} candidates and the ones with swapped mass assignments is fixed to the value extracted from simulation. No significant variation with $r$ was observed for the shape of the combinatorial background, or in the mean and in the root-mean-square of the distributions of signal \Dzero mesons or \Dzero candidates with swapped mass. Two examples of \PDz{} candidate invariant mass distributions, for \pp and \PbPb collisions, are available in \suppMaterial.

The raw \PDz{} radial distributions undergo several corrections, all calculated in bins of \ptd and $r$. First, the \Dzero meson yields are corrected for detector acceptance, and for trigger, track reconstruction, and selection inefficiencies. The correction factors are obtained from a \PYTHIA (\PYTHIA+\HYDJET) MC sample for the \pp (\PbPb) analysis. Second, the background contribution from combining a \Dzero meson with either a jet not coming from the same hard scattering or with a misreconstructed jet is subtracted using an event mixing technique, in which the background is estimated by combining the distributions of \Dzero-jet pairs formed with i) jets from the signal event and \Dzero mesons from minimum bias (MB) events~\cite{Khachatryan:2016odn}, ii) jets from MB events with \Dzero mesons from the signal event, and iii) jets and \Dzero mesons from MB events. In this procedure, each signal event is mixed with a MB event, which has a similar primary vertex position, amount of energy deposited in the forward hadronic calorimeters, and event plane angle~\cite{Chatrchyan:2012ta}. The resulting background radial distributions, which are in all cases less than 10\%, are then subtracted from the raw \Dzero radial distributions measured in the signal event. Finally, the background-subtracted radial distribution is corrected for jet resolution effects, using \PYTHIA+\HYDJET and \PYTHIA simulations, for the \PbPb and \pp results, respectively. The correction was calculated as the ratio between the \Dzero radial distributions after and before smearing the generated \ptj by energy and angular resolution corrections. 

Several sources of systematic uncertainty are considered for the \Dzero meson yield extraction and the jet reconstruction, and are studied in bins of \ptd and $r$. The uncertainty in the raw yield extraction (2.6--5.4\% for \pp and 1.4--8.2\% for \PbPb data) is evaluated by repeating the fit procedure using different background and signal fit functions and by varying the widths of the Gaussian functions that describe the \Dzero signal according to the differences (up to 20\%, as observed for the most forward region) between data and simulation. In the signal variation study, the sum of three Gaussian functions with the same mean but different widths is considered, while in the background variation study, a second-order polynomial function is used. This functional form gives a good description of the combinatorial background according to studies performed on same-sign pairs, which provide a pure combinatorial background with the same kinematic conditions. In these studies, the secondary vertex candidates are obtained by combining two same-sign tracks, which are assigned pion and kaon masses. The systematic uncertainty from the selection of the \PDz{} meson candidates (3.6 and 0.5\% for the low- and high-\ptd bin, respectively, for \pp, and 3.5 and 2.7\% for \PbPb data) is estimated by considering the differences in the \Dzero kinematic variables between simulation and data when applying each of the \Dzero candidate selection variables. The study is performed by varying one selection at a time and by considering the maximum relative discrepancies in the yield between data and simulation. The total uncertainty is the quadratic sum of the maximum relative discrepancy obtained by varying each of the three topological selection variables separately.

The systematic uncertainties for the jets include components for the uncertainty in the jet energy scale (JES) and jet energy resolution (JER). The systematic uncertainty pertaining to the JES is estimated by varying the \ptj by 2.8\% (in both \pp and \PbPb data), which represents the sum in quadrature of the observed data-to-simulation differences (2\%) and the nonclosure (\ie, deviation from unity) in simulation, when comparing reconstructed (detector-level) versus truth (generator-level) jets smeared by the known detector and reconstruction effects. An additional uncertainty 1.8--42\% for \PbPb data is added to account for the different detector response to quark versus gluon jets, since in \PbPb events, as opposed to \pp events, the quark- \vs gluon-initiated jet composition is not known because of the energy loss in the medium. The largest variation is observed at high \ptd and largest $r$ value, a region influenced by the small sample size. The assigned uncertainty represents the maximum difference from the nominal results when applying JES corrections obtained with a pure-gluon sample or a pure-quark sample.

The systematic uncertainty due to the JER in \PbPb collisions is estimated by varying the \ptj energy resolution by 15\% to account for an imperfect description of the fluctuations of the UE in the MC simulation. The variation considered is estimated by studying the effects of these fluctuations using two different methods: the random-cone technique \cite{Khachatryan:2016kdb,Khachatryan:2016jfl} and by embedding signal \PYTHIA dijet events into background \HYDJET samples. The random cone method consists of reconstructing many jets in a zero bias event, clustering particles in randomly placed cones in the entire $(\eta,\phi)$ space. When the method is applied in events with negligible contribution from hard scatterings, as is the case for zero bias events, the standard deviation of the distribution of \ptj obtained with this procedure can be used to estimate the magnitude of the UE fluctuations. The relative variations in the \Dzero spectra are 0.3--3.0\% in \pp and 0.6--5.6\% in \PbPb collisions. The systematic uncertainties from the trigger efficiency correction are estimated by the difference between the result with no correction and the nominal result, which are 0.3--2.7\% in \pp and 0.7--15\% in \PbPb data. Finally, a remaining nonclosure observed in MC between generated and reconstructed distributions of \Dzero mesons in jets, is corrected bin-by-bin. The magnitude of the correction is quoted as the systematic uncertainty in the resolution unfolding, which varies in the range 1.3--31\% in \pp and 0.7--32\% in \PbPb data. 

\begin{figure*}[htb]
\centering
\includegraphics[width= 0.49\textwidth]{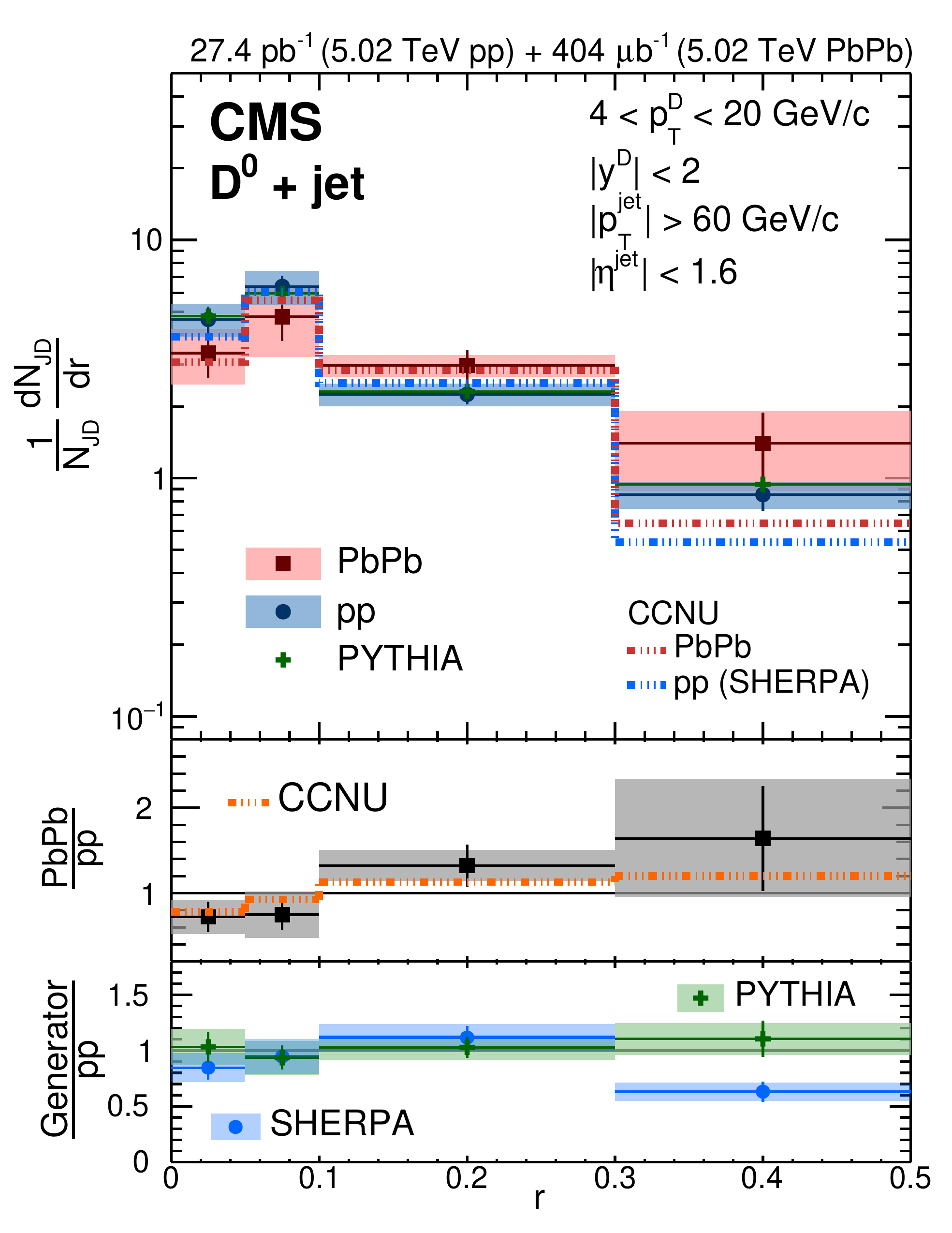}
\includegraphics[width= 0.49\textwidth]{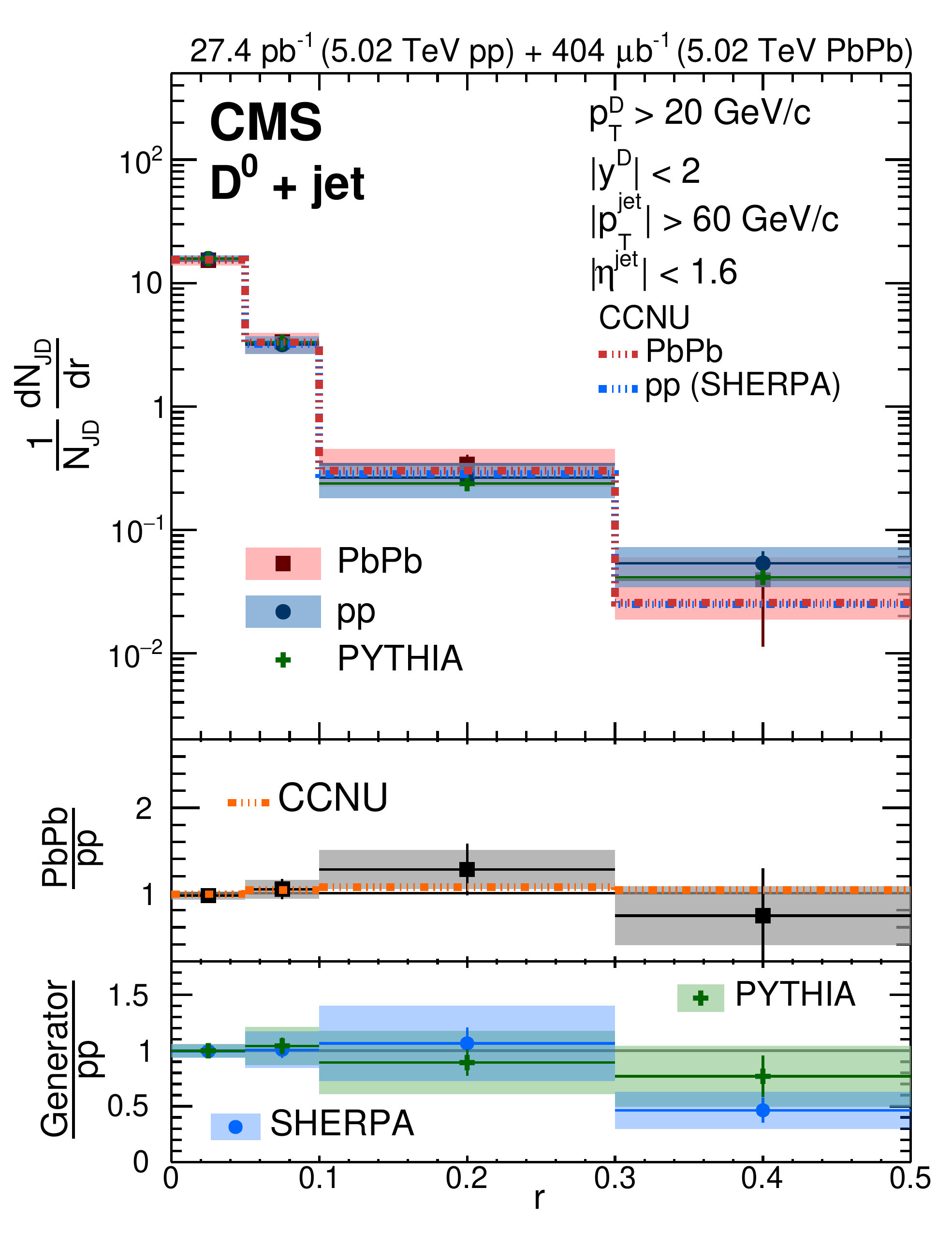}
\caption{Distributions of \Dzero mesons in jets, as a function of the distance from the jet axis ($r$) for jets of $\ptj>60\GeVc$ and $\abs{\etaj}<1.6$ measured in \pp and \PbPb collisions at $\sqrtsNN=5.02\TeV$. The measurement is performed in the \ptd range 4--20 \GeVc (left) and $\ptd>20\GeVc$ (right). Each spectrum is normalized to its integral in the region $0<r<0.3$. The vertical bars (boxes) correspond to statistical (systematic) uncertainties. The \PbPb spectra are compared to the CCNU energy loss model~\cite{Wang:2019xey}, while the \pp spectra are compared with predictions from the \PYTHIA and \SHERPA \pp MC event generators. The ratios of the \Dzero meson radial distributions in \PbPb and \pp data are shown in the middle panels. In the bottom panels the ratios of the \Dzero meson radial distributions of \pp over the two MC event generators are presented. }
\label{fig:cxsecsubdr}
\end{figure*}

The top panels of Fig.~\ref{fig:cxsecsubdr} show the measured \Dzero meson radial distributions in \pp and \PbPb collisions. The calculated $\langle r \rangle$ for the \PbPb (\pp) distributions is $0.198 \pm 0.015\stat \pm 0.005\syst$ ($0.160 \pm 0.007\stat \pm 0.009\syst$) and $0.048 \pm 0.002\stat \pm 0.004\syst$ ($0.046 \pm 0.001\stat \pm 0.003\syst$), for the low- and high-\ptd intervals, respectively. This result indicates that \Dzero mesons at low \pt are farther away from the jet axis in \PbPb compared to \pp collisions. At high \ptd, the measured spectra in \pp and \PbPb collisions fall rapidly, at a similar rate, as a function of $r$, similar to what was observed in inclusive jet-hadron correlation functions~\cite{Sirunyan:2018jqr}. 

The \pp results are compared to calculations from two \pp MC event generators: \PYTHIA~\cite{Sjostrand:2014zea}, a leading-order matrix element event generator, and \SHERPA~\cite{Gleisberg:2008ta}, which computes the next-to-leading QCD matrix elements matched to parton shower to generate the charm-jet events~\cite{Wang:2019xey}. For low-\pt \Dzero mesons, the measured spectrum in \pp collisions reaches a maximum at $0.05<r<0.1$, consistent with both \PYTHIA and \SHERPA~\cite{Wang:2019xey}. In the $r>0.3$ region however, \PYTHIA captures the features of the data better than \SHERPA, which underpredicts the \pp spectrum, in both \ptd intervals. The \PbPb   spectra is compared to an energy-loss model, CCNU~\cite{Wang:2019xey}, which uses \SHERPA for simulating the \pp baseline. The CCNU calculation includes in-medium elastic (collisional) and inelastic (radiative) interactions for both the heavy and the light quarks. This model, which predicts a small depletion (increase) of the \PDz{} meson yield at small (large) $r$ compared to \pp collisions, is consistent with data.  

To measure the medium modification of the radial profile, the ratio of \PbPb to \pp spectra is also presented in the first sub-panel of Fig.~\ref{fig:cxsecsubdr}. In this ratio, the uncertainties from JES, JER and \PDz{} candidate selections are considered uncorrelated between \pp and \PbPb datasets, and are not cancelled in the ratio. The uncertainties from the modeling of the signal shape, as well as the nonclosures observed, are partially cancelled: the systematic uncertainties are re-estimated directly on the ratio of the \PbPb to \pp yields. The ratio increases slightly as a function of $r$ at low \ptd, corresponding to a small shift of the \PDz{} mesons to larger radii in \PbPb, while the ratio is consistent with unity within the uncertainties at high \ptd. This shows that the modification of the radial profile of high \ptd is small. These features of the ratios at low and high \ptd are qualitatively different from inclusive charged particle radial distributions with respect to the jet axis measured in similar transverse momentum ranges~\cite{Sirunyan:2018jqr}. The inclusive measurements show a ratio significantly smaller than one, corresponding to a shift of the light quark mesons to smaller radii in \PbPb, for all tracks with $\pt>4\GeVc$ measured in jets with $\ptj>120\GeV$, for $r>0.1$ and more central \PbPb collisions. The CCNU model gives a good description of the ratio of \PbPb to \pp spectra. Although this ratio is less sensitive to the choice of \pp reference spectra, the \pp measurements presented in this Letter could improve the description of the \pp baseline.

In summary, this Letter presents the first measurement of the radial distributions of \Dzero mesons with respect to the jet axis in lead-lead (\PbPb) and proton-proton (\pp) collisions, performed with the CMS detector using jets with transverse momentum $\ptj> 60\GeVc$ and \Dzero mesons with $\ptd> 4\GeVc$. When compared to the results of Monte Carlo event generators, the radial distribution in \pp collisions is found to be well-described by \PYTHIA, while the slope of the distribution predicted by \SHERPA is steeper than that of the data. The modification of the \Dzero meson radial distributions in \PbPb collisions are studied by comparing them to those from \pp collisions. The comparisons hint at a modification of the \Dzero meson radial profile in \PbPb collisions at low \ptd that vanishes at higher \ptd. The results show that this modification is different from that of the light flavor hadrons. This measurement provides new experimental constraints on the mechanisms of heavy-flavor production in \pp collisions, as well as on the processes affecting the heavy quark propagation inside the quark-gluon plasma.

\begin{acknowledgments}
We congratulate our colleagues in the CERN accelerator departments for the excellent performance of the LHC and thank the technical and administrative staffs at CERN and at other CMS institutes for their contributions to the success of the CMS effort. In addition, we gratefully acknowledge the computing centers and personnel of the Worldwide LHC Computing Grid for delivering so effectively the computing infrastructure essential to our analyses. Finally, we acknowledge the enduring support for the construction and operation of the LHC and the CMS detector provided by the following funding agencies: BMBWF and FWF (Austria); FNRS and FWO (Belgium); CNPq, CAPES, FAPERJ, FAPERGS, and FAPESP (Brazil); MES (Bulgaria); CERN; CAS, MoST, and NSFC (China); COLCIENCIAS (Colombia); MSES and CSF (Croatia); RPF (Cyprus); SENESCYT (Ecuador); MoER, ERC IUT, PUT and ERDF (Estonia); Academy of Finland, MEC, and HIP (Finland); CEA and CNRS/IN2P3 (France); BMBF, DFG, and HGF (Germany); GSRT (Greece); NKFIA (Hungary); DAE and DST (India); IPM (Iran); SFI (Ireland); INFN (Italy); MSIP and NRF (Republic of Korea); MES (Latvia); LAS (Lithuania); MOE and UM (Malaysia); BUAP, CINVESTAV, CONACYT, LNS, SEP, and UASLP-FAI (Mexico); MOS (Montenegro); MBIE (New Zealand); PAEC (Pakistan); MSHE and NSC (Poland); FCT (Portugal); JINR (Dubna); MON, RosAtom, RAS, RFBR, and NRC KI (Russia); MESTD (Serbia); SEIDI, CPAN, PCTI, and FEDER (Spain); MOSTR (Sri Lanka); Swiss Funding Agencies (Switzerland); MST (Taipei); ThEPCenter, IPST, STAR, and NSTDA (Thailand); TUBITAK and TAEK (Turkey); NASU (Ukraine); STFC (United Kingdom); DOE and NSF (USA).
\end{acknowledgments}

\bibliography{auto_generated}

\ifthenelse{\boolean{cms@external}}{}{
\clearpage
\numberwithin{figure}{section}
\appendix
\section{Supplemental material\label{app:suppMat}}
\begin{figure*}[htb]
\centering
\includegraphics[width= 0.49\textwidth]{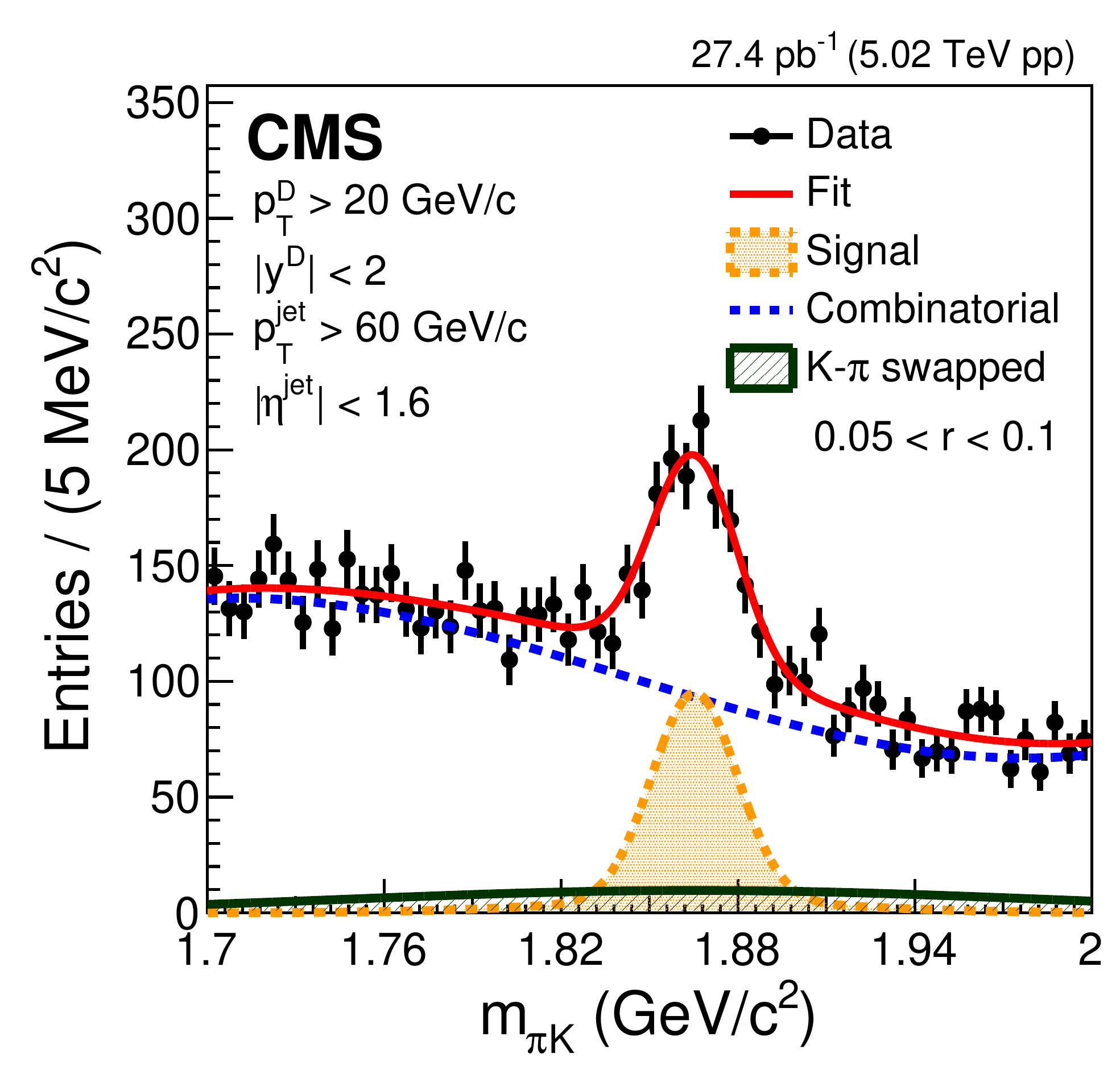}
\includegraphics[width= 0.49\textwidth]{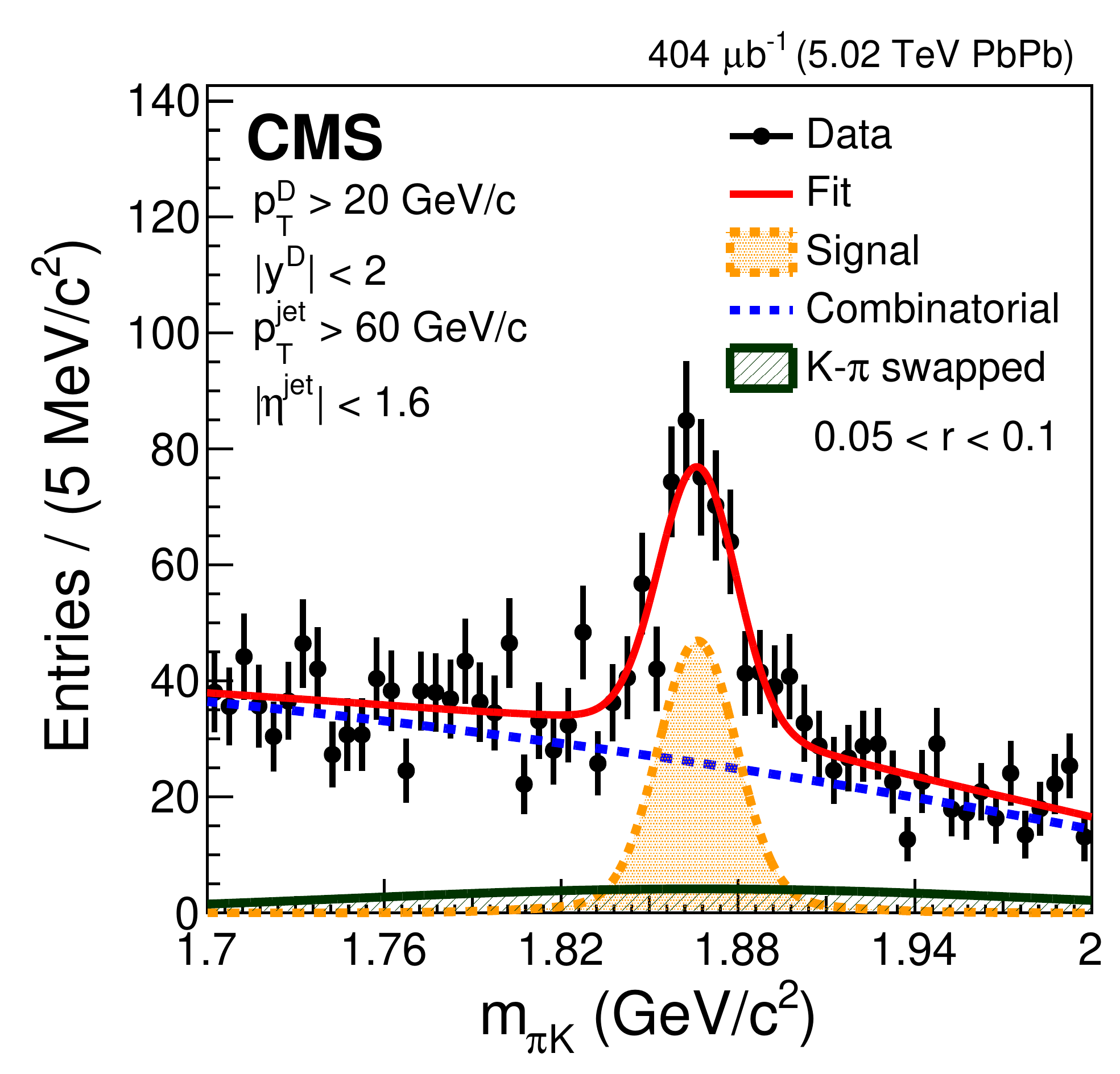}
\caption{Examples of raw \PDz{} candidate invariant mass distributions in \pp (left) and \PbPb (right) collisions at $\sqrtsNN=5.02\TeV$, for one $r$ interval, $0.05<r<0.1$.}
\label{fig:yieldsDzero}
\end{figure*}

}

\cleardoublepage \section{The CMS Collaboration \label{app:collab}}\begin{sloppypar}\hyphenpenalty=5000\widowpenalty=500\clubpenalty=5000\vskip\cmsinstskip
\textbf{Yerevan Physics Institute, Yerevan, Armenia}\\*[0pt]
A.M.~Sirunyan, A.~Tumasyan
\vskip\cmsinstskip
\textbf{Institut f\"{u}r Hochenergiephysik, Wien, Austria}\\*[0pt]
W.~Adam, F.~Ambrogi, E.~Asilar, T.~Bergauer, J.~Brandstetter, M.~Dragicevic, J.~Er\"{o}, A.~Escalante~Del~Valle, M.~Flechl, R.~Fr\"{u}hwirth\cmsAuthorMark{1}, V.M.~Ghete, J.~Hrubec, M.~Jeitler\cmsAuthorMark{1}, N.~Krammer, I.~Kr\"{a}tschmer, D.~Liko, T.~Madlener, I.~Mikulec, N.~Rad, H.~Rohringer, J.~Schieck\cmsAuthorMark{1}, R.~Sch\"{o}fbeck, M.~Spanring, D.~Spitzbart, A.~Taurok, W.~Waltenberger, J.~Wittmann, C.-E.~Wulz\cmsAuthorMark{1}, M.~Zarucki
% ...
\vskip\cmsinstskip
% ...
1:  Also at Vienna University of Technology, Vienna, Austria\\
% ...
\end{sloppypar}
\end{document}